\documentclass[aps,twocolumn,pra,showpacs,floatfix]{revtex4}
\usepackage{epsfig}
\usepackage{graphicx}
\usepackage{dcolumn}
\usepackage{amssymb,amsmath}
\usepackage{mathrsfs}
\begin{document}

\title{Time keeping and searching for new physics using metastable states of Cu, Ag and Au}

\author{V. A. Dzuba$^1$, Saleh O.  Allehabi$^1$,  V. V. Flambaum$^1$,  Jiguang Li$^2$ and S. Schiller$^3$}

\affiliation{$^1$School of Physics, University of New South Wales, Sydney 2052, Australia}
\affiliation{$^2$Institute of Applied Physics and Computational Mathematics, 6 Huayuan Road, Haidian District, Beijing, China}
\affiliation{$^3$Institut f\"{u}r Experimentalphysik, Heinrich-Heine-Universit\"{a}t D\"{u}sseldorf, 40225 
D\"{u}sseldorf, Germany}

\begin{abstract}
We study the prospects of using the electric quadrupole transitions from the ground states of Cu, Ag and Au to the metastable state $^2{\rm D}_{5/2}$  as clock transitions in optical lattice clocks.
We calculate lifetimes, transition rates, systematic shifts, and demonstrate that the 
fractional uncertainty of the clocks can be similar to what is achieved in the best current optical clocks. 
The use of these proposed clocks for the search of new physics, such as time variation of the fine structure constant, search for low-mass scalar dark matter,
violation of Local Position Invariance and violation of Lorenz Invariance is discussed. 
\end{abstract}

\maketitle

\section{Introduction}

Using optical clock transitions for searching for new physics beyond the standard model is a promising area of research.
A hypothetical manifestation of new physics at low energy is expected to be very small. Therefore, the highest possible accuracy
of the measurements is needed. Fractional uncertainty of the best optical clocks  
currently is around $1\times$
$10^{-18}$~\cite{Ludlow,Chou,Beloy1,Beloy2,Ushijima,Nicholson,Katori}, the highest accuracy 
so far achieved in the history of measurements.
However,  apart from a few exceptions (Hg$^+$, Yb$^+$) the best optical clocks are not 
sensitive to new physics such as time variation of the
fine structure constant, violation of Local Position Invarance (LPI) and violation of Local Lorentz Invariance (LLI),
etc.~\cite{qdot,EEP,Lia,Yb+LLI}. LPI, LLI and the Weak Equivalence Principle form the Einstein Equivalence Principle, the foundation of General Relativity.
Several ideas were proposed to combine high accuracy of optical clocks with high sensitivity to new physics.
These include the use of the highly charged ions (HCI)~\cite{H-likeHCI,HCIa,HCIb,HCIc}, nuclear clocks~\cite{th-clock},
and metastable atomic states with large value of the total angular momentum $J$ ($J > 1$)~\cite{qdot,Hg+clock0,Hg+clock,Yb-DFS,Yb+clock}. 
These states are connected to the ground state via transitions which correspond to single-electron transitions with large change of the single-electron
total angular momentum $j$. The large $\Delta j$ is what makes the transition to be sensitive to the variation of the fine structure constant (see e.g.~\cite{DFW99}). For example, in the present work we consider transitions between the $nd^{10}(n+1)s \ ^2$S$_{1/2}$ ground state and the $nd^{9}(n+1)s^2 \ ^2$D$_{5/2}$ excited metastable state. This is roughly the $s_{1/2}$ to $d_{5/2}$ transition with $\Delta j$=2.  

\begin{figure}[tb]
\epsfig{figure=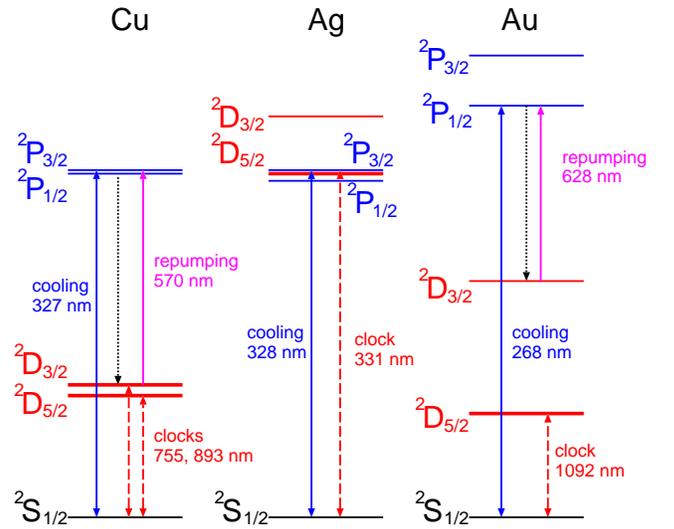,scale=0.6}
\caption{
	Energy diagram (approximately to scale) for the lowest states of Cu ($I=3/2$), Ag ($I=1/2$) and Au ($I=3/2$). 
	Thick red lines indicate the upper clock states. 
	 Electric quadrupole (E2) clock transitions are shown  as red dashed lines. Cooling transitions are shown as solid blue lines. The presence of leakage transitions (black dotted lines) implies the need for repumping (magenta lines). 
}
\label{f:EL}
\end{figure}

The energy diagrams for the five lowest states of Cu, Ag and Au studied in this work are presented in Fig.~\ref{f:EL}. The metastable state of interest ($^2$D$_{5/2}$) is the first excited state 
 for Cu and Au.
In Ag the $^2$P$^{\rm o}_{1/2}$ state lies below the $^2$D$_{5/2}$ clock state.
However, this has no significance since the states are very weakly connected (by the E3, M2 or hyperfine-induced E1 transitions). The clock transition in Ag was studied experimentally in Ref.~\cite{AgZ2}.

Sensitivity of the above metastable states to 
variation of the fine structure constant was studied before~\cite{qdot}. In this work we further study the states in terms of their suitability for high accuracy measurements and sensitivity to other manifestations of new physics, such as LPI violation and LLI violation. 

\section{Calculations}

We are mostly interested in five lowest states of Cu, Ag and Au shown on Fig.~\ref{f:EL}. Two out of the five states have excitations from a $d$-shell. This means that the $d$-shell is open and $d$-electrons should be treated as valence ones. The total number of valence electrons, eleven, is too large for most of standard calculational approaches. We use a version of the configuration interaction (CI) method specifically developed for such systems (the CIPT method~\cite{CIPT}). In this method off-diagonal matrix elements of the CI Hamiltonian between highly excited states are neglected. This allows to reduce the CI matrix to an effective  matrix of a small size in which contribution from high states is included perturbatively (see Ref.~\cite{CIPT} for details). 

We perform the calculations in the $V^{N-1}$ approximation, with one electron removed from initial relativistic Hartree-Fock (HF) calculations to obtain the potential for calculating single-electron basis states. 
Given the ground electronic configurations of the atoms here discussed, at first sight the best choice is the removal of the valence $s$ electron.
It turns out, however, that much better accuracy is achieved if a $d$ electron is removed instead. Thus, we perform the HF calculations for the [Ca]$3d^94s$ configuration of Cu, the [Sr]$4d^95s$ configuration of Ag, and the [Yb]$5d^96s$ configuration of Au.
The B-spline technique~\cite{B-spline}  is used to construct single-electron basis states above the core. Many-electron states for the CIPT calculations are constructed by exciting one or two electrons from a reference configuration and then using the resulting configurations to build all corresponding many-electron states of definite value of the total angular momentum $J$ and its projection $J_z$. States corresponding to about a hundred lowest non-relativistic configurations go into the 
effective CI matrix, while higher states are treated perturbatively. 
Note that our calculations are completely relativistic. We only use non-relativistic configurations to simplify the procedure of generating many-electron basis states. 
 In the list of non-relativistic configurations each of them is subsequently replaced by a corresponding set of relativistic configurations. 
E.g., the $5d^96s6p$ configuration is replaced by four relativistic ones, the $5d_{3/2}^45d_{5/2}^56s6p_{1/2}$,  $5d_{3/2}^35d_{5/2}^66s6p_{1/2}$, $5d_{3/2}^45d_{5/2}^56s6p_{3/2}$, and $5d_{3/2}^35d_{5/2}^66s6p_{3/2}$ configurations. 

To calculate transition amplitudes we use the well-known random phase approximation (RPA, see, e.g.~\cite{CPM}). The RPA
equations for  a single-electron state have the form
\begin{equation}\label{e:RPA}
(H^{\rm HF} - \epsilon_c)\delta \psi_c = -(\hat F + \delta V_{F}^{N-1})\psi_c.
\end{equation}
Here $H^{\rm HF}$ is the relativistic Hartree-Fock Hamiltonian, index $c$ numerates single-electron states, $\hat F$ is the operator of an external field, $\delta \psi_c$ is a correction to the state $c$ due to an external field, $\delta V_{F}^{N-1}$ is the correction to the self-consistent
Hartree-Fock potential due to the external field. The same $V^{N-1}$ potential is used in RPA and HF calculations. The RPA equations (\ref{e:RPA}) are solved self-consistently for all states $c$ in the core. Transition amplitudes are found  as matrix elements between many-electron states found in the CIPT calculations for the effective operator of an external field
\begin{equation}\label{e:Amp}
A_{ab} = \langle b| \hat F + \delta V^{\rm core}_{F} | a \rangle.
\end{equation}
Note that valence states are included in $V^{N-1}_F$ in (\ref{e:RPA}) but non included in the effective operator for valence states
$\delta V^{\rm core}_{F}$ in (\ref{e:Amp}).

The rates of spontaneous emission are given in atomic units by 
\begin{equation}\label{e:Td}
\Gamma_{\rm E1,M1} = \frac{4}{3}(\alpha\omega)^3 \frac{A^2_{\rm E1,M1}}{2J+1},
\end{equation}
for electric dipole (E1) and magnetic dipole (M1) transitions, and by 
\begin{equation}\label{e:Tq}
\Gamma_{\rm E2,M2} = \frac{1}{15}(\alpha\omega)^5 \frac{A^2_{\rm E2,M2}}{2J+1},
\end{equation}
for electric quadrupole (E2) and magnetic quadrupole (M1) transitions.
In these formulas, $\alpha$ is the fine structure constant, $\omega$ is the energy difference between the lower and upper states,
$A$ is the amplitude of the transition (\ref{e:Amp}), $J$ is the total angular momentum of the upper state.
The magnetic amplitudes $A_{\rm M1}$ and $A_{\rm M2}$ are proportional to the Bohr magneton $\mu_B =|e|\hbar/2mc$. Its numerical value in Gaussian-based atomic units is $\mu_B=\alpha/2 \approx 3.65 \times 10^{-3}$.
The lifetimes of excited states are calculated by $\tau_a =  2.4189 \times 10^{-17}/\sum_b \Gamma_{ab}$, where $\tau_a$ is the lifetime 
of atomic state $a$ in seconds, the summation goes over all possible transitions to lower states $b$, the transition probabilities $\Gamma_{ab}$ are given by (\ref{e:Td}) or (\ref{e:Tq}).

Energy levels and lifetimes of the five lowest states of Cu, Ag and Au are presented in Table~\ref{t:ECI}. 
Lifetimes were calculated using transition amplitudes and probabilities from Table~\ref{t:TR}. 
{Given the complexity of the considered systems the agreement between theory and experiment for the energies can be regarded as very good.
However,} the theoretical values for the lifetimes and transition rates are less accurate. This is partly due to the fact that the current version of the computer code includes perturbation correction to the energy but does not include a corresponding correction to the wave function~\cite{CIPT}. States with no excitation from the upper $d$-shell (e.g., the $5f^{10}6p_{1/2,3/2}$ states of Au) can be treated more accurately within different approaches, for example with the use of the correlation potential method~\cite{CPM}. The main advantage of the current approach is that it can be used for any state of considered atoms, including states with excitations from the upper $d$-shell where most of other methods would not work. 


	\begin{table}
		\caption{\label{t:ECI}
			Excitation energies ($E$, cm$^{-1}$) and lifetime for the lowest  states of Cu, Ag and  Au.}
		\begin{ruledtabular}
			\begin{tabular}{lllccc}
				&&
				\multicolumn{3}{r}{Energy [cm$^{-1}$]}&
				\multicolumn{1}{c}{Lifetime}\\
				
				\cline{4-5}
				\cline{6-6}
				
				\multicolumn{1}{c}{$N$}& 
				\multicolumn{1}{c}{Conf.}&
				\multicolumn{1}{c}{Term}&

				\multicolumn{1}{l}{NIST \cite{NIST}}&
				\multicolumn{1}{c}{Present work}&
				\multicolumn{1}{c}{Present work}\\
				
				\hline
					\multicolumn{6}{c}{Cu}\\

1 & $3d^{10}4s$&$^2${S}$_{1/2}$&0&0&$\infty$\\
2 &$3d^{9}4s^{2}$&$^2${D}$_{5/2}$&11203&10521&44.9 s\\
3&$3d^{9}4s^{2}$& $^2${D}$_{3/2}$ &13245&12270&7.3 s\\
4 & $3d^{10}4p$& $^2${P}$^{\rm o}$$_{1/2}$ &30535&29489&7.1 ns\\
5 & $3d^{10}4p$& $^2${P}$^{\rm o}$$_{3/2}$ &30784&31115&6.9 ns\\
				
					\multicolumn{6}{c}{Ag}\\
					
1 & $4d^{10}5s$& $^2${S}$_{1/2}$&0&0&$\infty$\\
2 &$4d^{10}5p$ & $^2${P}$^{\rm o}$$_{1/2}$ &29552&29495&6.6 ns\\
3&$4d^{10}5p$&$^2${P}$^{\rm o}$$_{3/2}$ &30473&30451&6.1 ns\\
4 & $4d^{9}5s^{2}$ & $^2${D}$_{5/2}$ &30242&32480&192 ms\\
5 & $4d^{9}5s^{2}$& $^2${D}$_{3/2}$ &34714&36430&174 $\mu$s\\
				
						\multicolumn{6}{c}{Au}\\
				1 & $5d^{10}6s$& $^2${S}$_{1/2}$&0&0&$\infty$\\
2 &$5d^{9}6s^{2}$ & $^2${D}$_{5/2}$ &9161&10671&43.7 s\\
3&$5d^{9}6s^{2}$ & $^2${D}$_{3/2}$ &21435&22096&33 ms\\
4 & $5d^{10}6p$& $^2${P}$^{\rm o}$$_{1/2}$ &37359&38853&4.1\footnotemark[1] ns\\
5 &$5d^{10}6p$& $^2${P}$^{\rm o}$$_{3/2}$&41175&43028&3.3\footnotemark[1] ns\\
				
				
			\end{tabular}		
			\footnotetext[1]{Experimental values are 6.0(1) ns for the $^2${P}$^{\rm o}_{1/2}$ state and 4.6(2) ns for the $^2${P}$^{\rm o}_{3/2}$ state \cite{Hannaford}.}
		\end{ruledtabular}
\end{table}
	
\begin{table}		
\caption{\label{t:TR} Transition amplitudes and probabilities for all possible transitions between the five lowest states of Cu, Ag and Au.}
\begin{ruledtabular}
\begin{tabular}{llrdc}

\multicolumn{1}{c}{Transition}& 
\multicolumn{1}{c}{Type}&
\multicolumn{1}{c}{$\hbar \omega$ [cm$^{-1}$]}&
				
\multicolumn{1}{c}{$|A|$ [a.u]}&
\multicolumn{1}{c}{$\Gamma$ [s$^{-1}$]}\\
								
\hline
\multicolumn{5}{c}{Cu}\\
				
2-1 & E2&11203&2.603&2.23$\cdot$10$^{-2}$\\
3-1 &M1&13245&0.0002\mu_B&6.27$\cdot$10$^{-7}$\\
4-1&E1& 30535 &2.217&1.42$\cdot$10$^{+8}$\\
5-1 & E1&30784&3.140&1.45$\cdot$10$^{+8}$\\
3-2 & M1&2043&1.549\mu_B&0.138\\
5-2 &E1&19581&0.546&1.13$\cdot$10$^{+6}$\\
4-3&E1& 17290&0.404&8.48$\cdot$10$^{+5}$\\
5-3 & E1&17538 &0.174&8.23$\cdot$10$^{+4}$\\
5-4 & M1&248&1.154\mu_B&1.38$\cdot$10$^{-4}$\\
				
\multicolumn{5}{c}{Ag}\\
				
2-1 & E1&29552&2.410&1.52$\cdot$10$^{+8}$\\
3-1 &E1&30473&3.382&1.64$\cdot$10$^{+8}$\\
4-1&E2& 30242 &3.325&5.22\\
5-1 & M1&34714&0.0002\mu_B&1.13$\cdot$10$^{-5}$\\
3-2 & M1&921&1.154\mu_B&7.01$\cdot$10$^{-3}$\\
5-2 &E1&5162&0.274&1.04$\cdot$10$^{+4}$\\
4-3&E1&230&0.368&0.839\footnotemark[1]\\
5-4 & M1&4472&1.549\mu_B&1.45\\
5-3 & E1&4242&0.119&5.45$\cdot$10$^{+2}$\\
				
\multicolumn{5}{c}{Au}\\

2-1 & E2&9161&4.359&2.29$\cdot$10$^{-2}$\\
3-1 &M1&21435&0.0008\mu_B&4.25$\cdot$10$^{-5}$\\
4-1&E1& 37359 &2.153&2.45$\cdot$10$^{+8}$\\
5-1 & E1&41175&2.923&3.02$\cdot$10$^{+8}$\\
3-2 & M1&12274&1.549\mu_B&29.9\\
5-2 &E1&32013&0.983&1.61$\cdot$10$^{+6}$\\
4-3&E1& 15924&0.504&1.04$\cdot$10$^{+6}$\\
5-3 & E1&19739 &0.243&2.30$\cdot$10$^{+5}$\\
5-4 & M1&3816&1.141\mu_B &0.488\\



\end{tabular}
\footnotetext[1]{Experimental value is 1.6(6) s$^{-1}$ \cite{Aglasercooling}.}
\end{ruledtabular}
\end{table}

\section{Analysis}

\subsection{Clock transitions}

Cu has two long-lived metastable states ($N=2,3$ in Tab.\,\ref{t:ECI}), Ag has one ($N=4$), and Au has one ($N=2$). Only the states of Cu and Au have lifetimes substantially larger than 1\,s, comparable to those of the currently used Sr and Yb lattice clocks. The 0.2\,s lifetime of the Ag upper clock state may turn out to be a competitive disadvantage for achieving top performance.
In the following we consider the clock transitions between these four
states and their respective ground states. {Note that the two clocks states of Cu are very similar and therefore in most of cases we present the data for only the $^2${D}$_{5/2}$ state.}

\subsection{Hyperfine structure}

The atoms considered here all exhibit hyperfine structure in the ground state, in  the clock state and in the excited state addressed in laser cooling. The nuclear spins are given in Tab.\,\ref{t:isotopes}. The hfs splitting is given by~\cite{Landau}
\begin{eqnarray}\label{e:hfs}
E_{\rm HFS}(F)&=&\frac{A}{2}F(F+1) + \frac{B}{2}\left[ F^2(F+1)^2 + \right.\\
&+&\left. F(F+1)[1-2J(J+1)-2I(I+1)]\right]. \nonumber 
\end{eqnarray}
The total angular momentum is $\mathbf{F}=\mathbf{J}+\mathbf{I}$, where $I$ is nuclear spin. $A$ and $B$ are magnetic dipole and electric quadrupole hfs constants, respectively.  They are reported in Tab.\,\ref{t:ABhfs}.

\begin{table}
	\caption{\label{t:isotopes}
		Stable isotopes with non-zero nuclear spin ($I$) and possible values of total angular momentum $F$ ($\mathbf{F}=\mathbf{I}+\mathbf{J}$) for ground and clock states of Cu, Ag, Au.
	}
	\begin{ruledtabular}
		\begin{tabular}{l cccc}
			\multicolumn{1}{c}{Isotopes}&
			\multicolumn{1}{c}{Transition}&
			\multicolumn{1}{c}{$I$}&
			\multicolumn{1}{c}{$F$ for GS} &
			\multicolumn{1}{c}{$F$ for CS} \\
			\hline
			$^{63,65}$Cu, $^{197}$Au & $^2$S$_{1/2} - ^2$D$_{5/2}$ & 3/2 & 1,2 & 1,2,3,4 \\
			$^{63,65}$Cu & $^2$S$_{1/2} - ^2$D$_{3/2}$ & 3/2 & 1,2 & 0,1,2,3 \\
			$^{107,109}$Ag & $^2$S$_{1/2} - ^2$D$_{5/2}$ &1/2 & 0,1 & 2,3 \\
		\end{tabular}
	\end{ruledtabular}
\end{table}

\begin{table}
	\caption{\label{t:ABhfs}
		Magnetic dipole ($A$) and electric quadrupole ($B$) hfs constants (MHz) used in the calculation of the second-order Zeeman shift.}
	\begin{ruledtabular}
		\begin{tabular}{r crc crr c}
			\multicolumn{1}{c}{Atom}&
			\multicolumn{3}{c}{Ground state}&
			\multicolumn{3}{c}{Clock state}&
			\multicolumn{1}{c}{Reference}\\
			&&\multicolumn{1}{c}{$A$}&
			\multicolumn{1}{c}{$B$} &&
			\multicolumn{1}{c}{$A$}&
			\multicolumn{1}{c}{$B$} & \\
			\hline
			$^{63}$Cu & $^2$S$_{1/2}$ &  5863 & 0 & $^2$D$_{5/2}$ &  749.1 & 186.0 & \cite{Cu-hfs,Cu-hfs1} \\
			$^{63}$Cu & $^2$S$_{1/2}$ &  5863 & 0 & $^2$D$_{3/2}$ &  1851.0 & 137.4 & \cite{Cu-hfs,Cu-hfs1} \\
			$^{107}$Ag &  $^2$S$_{1/2}$ & -1713 & 0 & $^2$D$_{5/2}$ &  -126 & 0 & \cite{AgZ2}\\
			$^{197}$Au & $^2$S$_{1/2}$ &   3050 & 0 & $^2$D$_{5/2}$ &  80.2 & -1049 & \cite{Au-hfs} \\
		\end{tabular}
	\end{ruledtabular}
\end{table}

For example, the hyperfine structure of Au has been studied experimentally with high precision in the 1960s, and has also been calculated \cite{Itano,Bieron}.
The hyperfine splitting between $F=1,2$ in the ground state amounts to 6.10\,GHz\,\cite{Dahmen}. 
The splittings in $^2$D$_{5/2}$ are  \cite{Au-hfs}

$F=1\leftrightarrow F=2$: 1.00\,GHz, 

$F=2\leftrightarrow F=3$: 0.71\,GHz, and 

$F=3\leftrightarrow F=4$: 0.52\,GHz. 

\noindent The splittings in $^2$D$_{3/2}$ are  \cite{Blachman} 

$F=0\leftrightarrow F=1$: 1.11\,GHz, 

$F=1\leftrightarrow F=2$: 1.31\,GHz, and

$F=2\leftrightarrow F=3$: 0.31\,GHz.  

\subsection{Laser cooling of Cu, Ag and Au}

\subsubsection{Silver}
Silver has been laser-cooled \cite{Aglasercooling}. Here, the cooling scheme is straightforward: the cooling transition is between ground and second excited state,  $^2$S$_{1/2}\rightarrow^2$P$_{3/2}$, so that there are no leakage channels to other electronic states. No repumper laser fields are needed.

\subsubsection{Gold}
\begin{figure}[tb]
\epsfig{figure=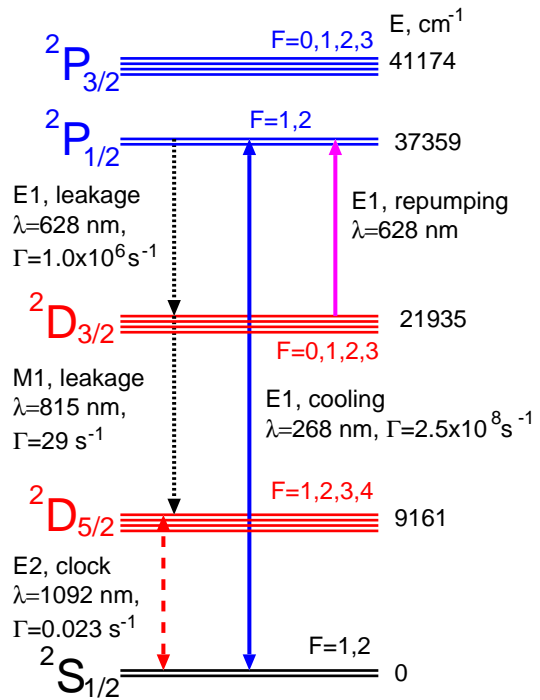,scale=0.9}
\caption{Details of the level scheme of $^{197}{\rm Au}$ ($I=3/2$) (not to scale) with proposed laser cooling. The hyperfine structure is shown schematically. The magenta arrow shows the repumper transition. Narrow-linewidth laser cooling is not shown. The clock transition (dashed red line) is composed of several hyperfine components.}
\label{f:Au-cool}
\end{figure}

A scheme for laser cooling of Au is presented in Fig.~\ref{f:Au-cool}. The main cooling transition is the electric dipole transition between the ground state and the excited odd-parity $^2$P$^{\rm o}_{1/2}$ state. Compared to using $^2$P$^{\rm o}_{3/2}$ as upper level, the advantage is that only one repumper is needed and that the longer cooling wavelength is experimentally advantageous. There is leakage from the $^2$P$^{\rm o}_{1/2}$ to the $^2$D$_{3/2}$ state by another electric dipole transition ($4\rightarrow3$). Therefore, without repumping only $\sim$~250 cycles 
are possible. With repumping (628\,nm) the cooling may go for as long as needed. Another leakage channel is too weak to affect the scheme. 

\subsubsection{Copper}
A cooling scheme similar to silver can be considered for copper: 
$^2$S$_{1/2}(F_g=2)\rightarrow ^2$P$_{3/2}(F_e=3)$. 

\subsubsection{Additional remarks}

Optical lattice clocks require the cooling of atoms to the $\mu$K level for efficient loading of the optical lattice with the atoms.
Therefore, after cooling on the strong E1 transition to a temperature on the order of 1\,mK, a second cooling process utilizing a weak transition should follow ("narrow-linewidth cooling"). 
One option is to cool on the $^2$S$_{1/2}\rightarrow$ $^2$D$_{3/2}$ transition ($1-3$ for Cu, $1-5$ for Ag, $1-3$ for Au). These are M1 transitions and are very weak. 
However, the strengths could be increased and the lifetime of the $^2$D$_{3/2}$ states shortened by E1 coupling them to the respective $^2$P states using appropriate waves. 

The hyperfine structure in both lower and upper laser cooling levels will typically require additional repumper fields to optimize cooling efficiency. We shall not discuss such experimental details here.

Finally, we note that copper and silver atoms have been cooled using buffer-gas cooling \cite{CuAg-cool}. 

\subsection{Polarizabilities, black-body radiation shifts and magic frequencies}

Knowledge of the atomic polarizabilities for both states of the clock transition is important for estimation of the frequency shift caused by black-body radiation and for finding the so called {\it magic} frequency of the lattice laser field, i.e. the frequency at which the dynamic polarizabilities of both states are equal, causing no frequency shift.

The static scalar polarizability $\alpha_v(0)$ of an atom in state $v$ is given by
\begin{equation}\label{e:pol0}
\alpha_v(0) = \frac{2}{3(2J_v+1)}\sum_n\frac{|\langle v||D||n\rangle |^2}{E_n-E_v},
\end{equation}
where $D$ is the electric dipole operator with the RPA correction (see the previous section), and the summation goes over the complete set of excited many-electron states. 

 Static scalar polarizabilities of the ground states of Cu, Ag and Au are known from a number of calculations and measurements~\cite{pol0}. Table~\ref{t:pol} presents the recommended values taken from Ref.~\cite{pol0}. In contrast, to the best of our knowledge, there is no similar data for the upper clock states of Cu, Ag and Au. Therefore, we performed the calculations using two different approaches. 
 
 In the first approach we stay within the CIPT method and calculate twenty odd-parity states for each value of the total angular momentum $J$ which satisfies the electric dipole selection rules for the transitions from the ground and clock states ($J=1/2, 3/2, 5/2, 7/2$). Then we use the formula (\ref{e:pol0}) to perform the calculations for both states. These calculations show three important things: (a) there is good agreement with other data for the ground state, (b) there is good saturation of the summation in (\ref{e:pol0}), (c) the summation for the clock states is strongly dominated by the transitions to the states of the $5d^96s6p$ configuration (we use the Au atom as an example). 
 
The last fact implies that a different approach can be used, previously suggested for atoms with open $f$-shells~\cite{AKozlov}. In this second approach we use the fact that the sum (\ref{e:pol0}) is dominated by the $6s$ - $6p$ transitions while the open $5d^9$ subshell remains unchanged.  Therefore, the open $d$-shell is attributed to the core and treated as a closed shell with an occupational number of 0.9. The atom is treated as a system with two external electrons above the closed-shell core and an appropriate CI+MBPT~\cite{CI+MBPT} method is used (see Ref.~\cite{AKozlov} for more details). The advantage of this approach is the efficient completeness of the basis with two electron excitations. The shortcoming is the omission of the transition amplitudes involving excitations from the $d$-shell. In contrast, the CIPT approach includes all amplitudes; however the summation in (\ref{e:pol0}) is truncated much earlier. 
 
 In the end, both approaches give similar results. 
 The results for the clock states are presented in Table~\ref{t:pol} together with estimated uncertainties. For these estimations we used a comparison of the two approaches for the clock states as well as comparison of the CIPT calculation with other data for the ground states.

The results of the calculations indicate that the values of the polarizabilities of the clock states of Cu, Ag and Au are similar to those of the ground state. This is a non-standard situation. More often, the polarizabilities of excited states are larger. Indeed, the higher is the  state on the energy scale, the smaller is the energy denominator in (\ref{e:pol0}). Present results can be explained by the fact that summation in (\ref{e:pol0}) is dominated by the states of the $5d^{10}np$ configurations for the ground state (we use Au again as an example) and by the states of the $5d^96s6p$ configuration for the clock state. The later states are higher on the energy scale. 

The BBR shift  is given by (see, e.g.~\cite{BBR})
\begin{equation}\label{e:BBR}
\delta \nu_{\rm BBR} = -\frac{2}{15}(\alpha \pi)^3 T^4 \left(\alpha_c(0) - \alpha_g(0)\right),
\end{equation}
where $\alpha$ is the fine structure constant, $T$ is temperature, $\alpha_c(0)$ and  $\alpha_g(0)$ are static scalar polarizabilities of the clock and ground states. For simplicity, we do not include the dynamic correction to the BBR shift. For the more complete formula see, e.g.~\cite{BBR}.

The BBR shift is proportional to the difference between the polarizabilities of the two states. The similarity of the polarizabilities implies a substantial cancellation of the black-body radiation (BBR) frequency shift, a very favourable effect.  
Since the differences are smaller or close to the uncertainty of both numbers we can only give upper limits for the BBR shifts. The results are presented in Table~\ref{t:pol}. These limits are lower than the shifts in the standard ytterbium and strontium lattice clocks. More accurate estimations might be possible if the polarizabilities are measured or calculated to higher accuracy.

Magic frequencies can be found in the vicinity of every resonance for one of the polarizabilities, i.e. when the frequency of the lattice laser field is approximately equal to the excitation energy (energy denominator in (\ref{e:pol0})). The first magic frequency is near the first resonance for the ground state polarizability, i.e. $\hbar \omega_m \simeq  30535\,{\rm cm}^{-1}$ for Cu, $\hbar \omega_m \simeq  29552\,{\rm cm}^{-1}$ for Ag and $\hbar \omega_m \simeq  37359\,{\rm cm}^{-1}$ for Au. Note that since the clock states have large values of the total angular momentum ($J=5/2$), the magic frequencies would also depend on the quadrupole contribution to the polarizabilities. The current level of computational accuracy does not allow to find accurate values of the magic frequencies. Having more experimental data may help. In the vicinity of a resonance or a few resonances a semi-empirical formula can be used
\begin{equation}\label{e:pw}
\alpha_a(\omega) \approx \alpha^{\prime}_a(0) + \frac{2}{3(2J_v+1)}\sum_b \frac{A_{ab}^2}{\omega - \Delta E_{ab}},
\end{equation}
where $\alpha^{\prime}_a(0)$ is chosen in such a way that $\alpha_a(\omega=0)$ is equal to known (e.g. experimental)
static polarizability of state $a$. Summation in (\ref{e:pw}) goes over close resonances. If the static polarizability is known to sufficient accuracy and amplitudes $A_{ab}$ of E1-transitions are extracted from experimental data or from accurate atomic calculations, then (\ref{e:pw}) can be used to find magic frequencies.  

\begin{table*}
  \caption{\label{t:pol}
  	Scalar static polarizabilities (in $a_B^3$) and BBR frequency shift for clock 
  	transitions of Cu, Ag and Au. 
	$\Delta\alpha$ is the difference between the theoretical value for the upper clock state and the experimental value of the lower clock state.} 
\begin{ruledtabular}
\begin{tabular}{l c c|cc c|c|cc}
\multicolumn{1}{c}{Atom}&
\multicolumn{2}{c|}{$\alpha_g(0)$}&
\multicolumn{3}{c|}{$\alpha_c(0)$}&
\multicolumn{1}{c|}{$\Delta\alpha$}&
\multicolumn{2}{c}{BBR ($T=300\,$K)} \\
& \multicolumn{1}{c}{Expt.~\cite{pol0}} &
\multicolumn{1}{c|}{CIPT} &
\multicolumn{1}{c}{CIPT} &
\multicolumn{1}{c}{CI+MBPT} & 
\multicolumn{1}{c|}{Final} &&
\multicolumn{1}{c}{$\Delta\nu$\,[Hz]}&
\multicolumn{1}{c}{$\Delta\nu/\nu$} \\
\hline
Cu\tablenotemark[1]     &  47(1) & 54.5 & 46.8 & 42.9 & 45(8) & 2(8) & $ <0.09$ & $<2.6 \cdot 10^{-16}$ \\
Ag     &  55(8) & 51.8 & 45.9 & 49.5 & 47(2) & -8(8) & $ <0.14$ & $<1.5 \cdot 10^{-17}$ \\
Au     &  36(3) & 35.7 & 38.9 & 33.2 & 36(3) &  0(4) & $ <0.03$ & $<5.6 \cdot 10^{-17}$ \\
\end{tabular}
\tablenotetext[1]{State $c$ is the $^2$D$_{5/2}$ clock state.}
\end{ruledtabular}
\end{table*}

\subsection{Stark, quadrupole and Zeeman shifts}


Interaction of atomic electrons with external electric field and its gradient lead to Stark and electric quadrupole shifts of transition frequencies.
These shifts are tiny in optical lattice clocks~\footnote{On the inner surface of a metallic vacuum chamber there can be spatial  variations of the electrostatic potential of order 0.1 V. The typical size of a vacuum chamber may be 10 cm. Thus, the electric field gradient is smaller than 0.1 V/(10 cm$)^2$. Assuming a typical quadrupole moment $Q \sim 1$~a.u. leads to a negligible quadrupole shift $\sim 10^{-5}$~Hz. The corresponding Stark shift is $\sim 10^{-7}$~Hz.}. We consider the shifts 
in more details in the Appendix.



The linear Zeeman shift is given by the expression 
\begin{equation}\label{e:z1}
\Delta E_{F,F_z} = g_F \mu_B B F_z,
\end{equation}
where $g_F$ is the $g$-factor of a particular hfs state. It is related to the electron $g_J$-factor by
\begin{equation}\label{e:f}
g_F=g_J \langle F,F_z=F,I,J|\hat J_z|F,F_z=F,I,J \rangle/F.
\end{equation}
Electron $g$-factors have approximate values $g_{1/2} \approx 2$, $g_{3/2} \approx 0.8$, $g_{5/2} \approx 1.2$. More accurate values for Cu, Ag and Au can be found in NIST tables~\cite{NIST}. 
For a clock state with $J=5/2$ and $F=2$ we have $g_2= (11/12)g_{5/2}=1.1$.
For a clock state with $J=3/2$ and $F=2$ we have $g_2= (1/2)g_{3/2}=0.4$.

The linear Zeeman shift can be avoided if only transitions between states with $F_z = 0$ are considered, as sug-
gested in the past for clock operation.
Alternatively, one can average over the transition frequencies with positive and negative $F_z$ in order to cancel the linear shift. However, the large individual shifts will make it difficult to achieve an accurate cancellation.   

A second-order Zeeman shift is unavoidable. Therefore, it is important to know its value. If we consider transitions between definite hfs components, then the shift is strongly dominated by transitions within the same hfs multiplet. The total shift is the difference between the second-order shifts in the clock and in the ground state. Both shifts are given by 
\begin{eqnarray}\label{e:z2}
&&\delta E_{F,F_z} = \\
&&\sum_{F'=F \pm 1,F'_z}\frac{|\langle F'F'_zIJ|\hat J_z|FF_zIJ \rangle g_J \mu_B B |^2}{\Delta E_{\rm{HFS}}(F,F')}. \nonumber
\end{eqnarray} 
Here $\Delta E_{\rm{HFS}}(F,F')=E(F\,I\,J)-E(F'\,I\,J)$ is the hfs interval. 
It has a different sign depending on whether this is an up or down transition. 

It follows from (\ref{e:hfs}) that 
\begin{eqnarray}
&& \Delta E_{\rm HFS}(F,F+1) = -A(F+1) -  \nonumber \\
&&B \left( 2(F+1)^2 +1-2J(J+1)-2I(I+1)\right), \nonumber 
\end{eqnarray}
and
\begin{eqnarray}
 &&\Delta E_{\rm HFS}(F,F-1)  = \nonumber \\
 &&AF + B \left( 2F^2 +1-2J(J+1)-2I(I+1)\right).\nonumber
 \end{eqnarray}
  Using experimental values for $A$ and $B$ (see Table~\ref{t:ABhfs}) we calculate the second-order Zeeman shift for Cu, Ag and Au. The results are presented in Tables~\ref{t:AgZ2} and \ref{t:CuAuZ2}.
The shift for Ag was studied before~\cite{AgZ2}. 
Our result differs from theirs; this may be due to a simple calculational error.
Table~\ref{t:AgZ2} presents separate contributions from the shifts in the ground and excited states. One can see that the disagreement may come from the sing error in a particular contribution. Different signs are caused by energy denominators.
For example, when we move from first to the second line of the table, the sign of the energy denominator for the ground state contribution changes and so does the contribution itself. Since other contribution remains the same, the total shift must change.

Table~\ref{t:CuAuZ2} shows the second-order Zeeman shift for $^{63}$Cu and $^{197}$Au. As in the case of $^{107}$Ag the shift is small.
Note that Cu has one clock transition with both tiny quadratic shift coefficient and no linear shift.
By measuring two or more $F_{gz}=0\rightarrow F_{ez}=0$ Zeeman components and taking appropriate combinations of the corresponding transition frequencies the second-order shift may be substantially reduced.

{The quadratic shift vanishes in the considered approximation for transitions between states with maximum value of $F$ and its projection $F_z$ (see bottom lines of Table~\ref{t:CuAuZ2}). This is because there are no terms in (\ref{e:z2}) which would satisfy the selection rules. Note also that the (non-zero) numbers in Table~\ref{t:CuAuZ2} should be considered as rough estimations only. This is because of uncertainties of the experimental data for the electric quadrupole hfs constant $B$, in particular for Cu~\cite{Cu-hfs1}. The numbers can change several times depending on which set of data is used.}

\begin{table}
  \caption{\label{t:AgZ2}
  Second-order Zeeman shift (mHz/$(\mu T)^2$) for $^{107}$Ag and comparison with other calculations.
  Index $g$ is for the ground state, index $c$ is for the excited (clock) state. It is assumed that $F_z=0$.}
\begin{ruledtabular}
\begin{tabular}{cc dddd}
\multicolumn{1}{c}{$F_c$}&
\multicolumn{1}{c}{$F_g$}&
\multicolumn{1}{c}{$\Delta E_c$}&
\multicolumn{1}{c}{$\Delta E_g$}&
\multicolumn{1}{c}{$\Delta E_c-\Delta E_g$}&
\multicolumn{1}{c}{Ref.~\cite{AgZ2}}\\
\hline
2 & 0 & 0.186 & 0.114 & 0.072 & 0.07 \\
2 & 1 & 0.186 & -0.114 & 0.301 & 0.07 \\
3 & 1 & -0.186 & -0.114 & -0.072 & -0.3 \\
\end{tabular}
\end{ruledtabular}
\end{table}

\begin{table}
  \caption{\label{t:CuAuZ2}
  Second-order Zeeman shift (mHz/$(\mu T)^2$) for $^{63}$Cu  and $^{197}$Au. Gaps in the data mean that corresponding set of quantum numbers is not possible for the transition.}
\begin{ruledtabular}
\begin{tabular}{cccc ddd}
\multicolumn{1}{c}{$F_g$}&
\multicolumn{1}{c}{$F_{gz}$}&
\multicolumn{1}{c}{$F_c$}&
\multicolumn{1}{c}{$F_{cz}$}&
\multicolumn{3}{c}{$\Delta E_c-\Delta E_g$}\\
&&&&\multicolumn{1}{c}{$^{63}$Cu}&
\multicolumn{1}{c}{$^{63}$Cu}&
\multicolumn{1}{c}{$^{197}$Au}\\
&&&&\multicolumn{1}{c}{$^2$D$_{5/2}$}&
\multicolumn{1}{c}{$^2$D$_{3/2}$}&
\multicolumn{1}{c}{$^2$D$_{5/2}$}\\
\hline
1 & 0 & 0 & 0 &           & -0.759 & \\
1 & 0 & 1 & 0 &  0.087 & 0.743  & 0.023 \\
1 & 0 & 2 & 0 &  0.193 & 0.058 & 0.027 \\
1 & 0 & 3 & 0 &  -0.247 & 0.025 & 0.050 \\
2 & 0 & 0 & 0 &              & -0.792 &  \\
2 & 0 & 1 & 0 & 0.053 & 0.710 & -0.041 \\
2 & 0 & 2 & 0 & 0.160 & 0.024 & -0.037 \\
2 & 0 & 3 & 0 & -0.281 & -0.009 & -0.014 \\
2 & 0 & 4 & 0 &  0.001 & & -0.037 \\
2 & $\pm2$ & 3 & $\pm2$ & -0.044 & 0.004 & 0.002 \\
2 & $\pm2$ & 3 & $\pm3$ & -0.017 & 0.0 & 0.004 \\
2 & $\pm2$ & 4 & $\pm4$&  0.0 &  & 0.0 \\
\end{tabular}
\end{ruledtabular}
\end{table}

\section{Search for new physics}

An exceptionally high accuracy of atomic clocks is a great advantage for using them  in a search for new physics. The search is conducted by
monitoring relative values of different atomic frequencies over a significant time interval. Hypothetical time-variation of the frequency ratio allows multiple interpretations. 
E.g., the interaction between low-mass scalar dark matter and ordinary matter may lead to
oscillation  of the fine structure constant and transient variation effect~\cite{DM1,DM2,DM3}. 
In this section we consider time variation of the fine structure constant $\alpha$ ($\alpha = e^2/\hbar c$),
Local Position Invariance (LPI) violation and Local Lorentz Invariance (LLI) violation. 

\subsection{Time variation of the fine structure constant}

It is convenient to present to parametrise the $\alpha$-dependence of atomic frequencies by the formula $\omega = \omega_0 + q[(\frac{\alpha}{\alpha_0})^2-1]$~\cite{CJP}, where $\alpha_0$ and $\omega_0$ are present-day values of the fine structure constant and the frequency of the transition, $q$ is the sensitivity coefficient which comes from the calculations. To monitor possible frequency change one 
atomic frequency is measured against another. Then
\begin{equation}\label{e:w1w2}
\frac{\partial}{\partial t}\ln \frac{\omega_1}{\omega_2} = \frac{\dot \omega_1}{\omega_1} -  \frac{\dot \omega_2}{\omega_2} =
\left(\frac{2q_1}{\omega_1} - \frac{2q_2}{\omega_2}\right) \frac{\dot \alpha}{\alpha}.
\end{equation}
The value $K=2q/\omega$ 
is called an enhancement factor. It shows that if $\alpha$ changes in time then 
$\omega$ changes $K$ times faster.
Calculated values of $q$ and $K$ for different optical clock transitions are presented in Table~\ref{t:q}. Note that we include not all known clock
transitions but only those which are sensitive for $\alpha$-variation searches. There are six transitions where $|K|>1$. The largest values of $|K|$ correspond to the smallest values of transition frequency $\omega$. It would be wrong to say that all these transitions are good for searching for $\alpha$-variation. This is because the accuracy of the measurements is equally important (see also discussion in Ref.~\cite{Yb-DFS}).
The true figure of merit is the  ratio of the relative frequency shift due to variation of  $\alpha$ and the 
fractional uncertainty of the measurements,
$(q/\omega)/(\delta \omega/\omega) = q/\delta \omega$. This ratio does not depend on $\omega$. Therefore, looking for a large value of $K$ caused by the small value of $\omega$ brings no benefit. The value of the relativistic energy shift $q$ is more important. Comparing the values of $q$ for different clock transitions (see Table~\ref{t:q}) we see that the E2 clock transition in Au 
is essentially as good as the recently proposed new transitions in neutral ytterbium and only 30\% smaller than the octupole transition in the ytterbium ion (Yb\,II).

\begin{table*}
  \caption{\label{t:q}
    Sensitivity of clock transitions to variation of the fine structure constant ($q, K$), 
    to LLI violation (reduced matrix element $\langle c||T_0^{(2)}||c\rangle$ of the tensor operator (\ref{e:LLI}) for the upper state $c$), 
    and to LPI violation (relativistic factor $R$). Note that  $\langle g||T_0^{(2)}||g\rangle$ is zero for the ground state due to the small value of the total angular momentum $J=1/2$.}
\begin{ruledtabular}
\begin{tabular}{l cccdddd dd}
\multicolumn{1}{c}{Atom/}&
\multicolumn{3}{c}{Transition}&
\multicolumn{1}{c}{$\hbar \omega$\footnotemark[1]}&
\multicolumn{1}{c}{$q$ }&
\multicolumn{1}{c}{$K=2q/\hbar \omega$}&
\multicolumn{1}{c}{$\langle c||T_0^{(2)}||c \rangle$}&
\multicolumn{2}{c}{$R$}\\
\multicolumn{1}{c}{Ion}&
\multicolumn{1}{c}{Lower state}&&
\multicolumn{1}{c}{Upper state}&
\multicolumn{1}{c}{(cm$^{-1}$)}&
\multicolumn{1}{c}{(cm$^{-1}$)}&&
\multicolumn{1}{c}{(a.u.)}& 
\multicolumn{1}{c}{Present}&
\multicolumn{1}{c}{Other}\\
\hline
Cu     &  $3d^{10}4s \ ^2$S$_{1/2}$ &-& $3d^{9}4s^2 \ ^2$D$_{5/2}$ & 11202.565 & -4000 & -0.71 &-48 & 0.98 & \\
Cu     &  $3d^{10}4s \ ^2$S$_{1/2}$ &-& $3d^{9}4s^2 \ ^2$D$_{3/2}$ & 13245.443 & -1900 & -0.29 &-37 & 0.99 & \\
Ag     &  $4d^{10}5s \ ^2$S$_{1/2}$ &-& $4d^{9}5s^2 \ ^2$D$_{5/2}$ & 30242.061 & -11300 & -0.75  & -41 & 0.93 & \\
Au     &  $5d^{10}6s \ ^2$S$_{1/2}$ &-& $5d^{9}6s^2 \ ^2$D$_{5/2}$ & 9161.177 & -38550 & -8.4  & -45 & 0.67 & \\
Hg~II\footnotemark[2] &  $5d^{10}6s \ ^2$S$_{1/2}$ &-& $5d^{9}6s^2 \ ^2$D$_{5/2}$ & 35514.624 & -52200 & -2.94 && 0.68 & 0.2\footnotemark[3] \\
Yb\footnotemark[4]    &  $4f^{14}6s^2 \ ^1$S$_{0}$ &-& $4f^{14}6s6p \ ^3$P$^{\rm o}_{0}$ & 17288.439 & 2714 & 0.31 &&  1.12 & 1.20\footnotemark[3] \\
Yb\footnotemark[4]    &  $4f^{14}6s^2 \ ^1$S$_{0}$ &-& $4f^{13}5d6s^2 J=2$ & 23188.518 & -44290 & -3.82&& 0.65 & 1.40\footnotemark[4] \\
Yb\footnotemark[5]     &  $4f^{14}6s6p \ ^1$P$^{\rm o}_{0}$ &-& $4f^{13}5d6s^2 J=2$ & 5900.079 & -43530 & -15 && & \\
Yb~II\footnotemark[6]  &  $4f^{14}6s \ ^2$S$_{1/2}$ &-& $4f^{13}6s^2 \ ^2$F$^{\rm o}_{7/2} $ & 21418.75 & -56737 & -5.3  &-135& 0.58 & -1.9\footnotemark[3] \\
Yb~II\footnotemark[6]  &  $4f^{14}6s \ ^2$S$_{1/2}$ &-& $4f^{14}5d \ ^2$D$_{3/2} $ & 22960.80 & 10118 & 0.88  &10&1.42& 1.48\footnotemark[3] \\
Yb~II\footnotemark[6]  &  $4f^{13}6s^2 \ ^2$F$^{\rm o}_{7/2} $ &-& $4f^{14}5d \ ^2$D$_{3/2} $ & 1542.06 & -66855 & -87 &&&\\
\end{tabular}
\footnotetext[1]{NIST~\cite{NIST}.}
\footnotetext[2]{Ref.~\cite{CJP,Yb-DFS}.}
\footnotetext[3]{Ref.~\cite{EEP}.}
\footnotetext[4]{Ref.~\cite{Yb-DFS}.}
\footnotetext[5]{Ref.~\cite{SafronovaYb}.}
\footnotetext[6]{Ref.~\cite{Yb+LLI,YbIIq}.}
\end{ruledtabular}
\end{table*}

\subsection{LPI violation}

In the standard model extension, the term in the Hamiltonian responsible for the LPI violation can be presented in the form
(see, e.g., Ref.~\cite{EEP})
\begin{equation}\label{e:EEP}
\hat H_{\rm EEP} = c_{00}\frac{2}{3}\frac{U}{c^2}\hat K,
\end{equation}
where $c_{00}$ is the unknown parameter characterizing the magnitude of the LPI violation, $U$ is the gravitational potential, $c$ 
is the speed of light, $\hat K= c\gamma_0\gamma^j p_j/2$ is the relativistic operator of kinetic energy in which $\gamma_0$ and 
$\gamma^j$ are Dirac matrices, and $\mathbf{p}=-i\hbar \mathbf{\nabla}$ is electron momentum operator.

The presence of the term (\ref{e:EEP}) in the Hamiltonian would manifest itself via a dependence of the atomic frequencies on the time in the year,
caused by the changing Sun-Earth distance leading to change of the Sun's gravitational potential $U$. As in the case of the $\alpha$-variation, at least two clock transitions are needed to measure one clock frequency against the other.
The interpretation of the measurements is based on the formula~\cite{EEP}
\begin{equation}\label{e:REEP}
\frac{\Delta \omega_1}{\omega_1} - \frac{\Delta \omega_2}{\omega_2} = - (R_1 - R_2)\frac{2}{3}c_{00}\frac{\Delta U}{c^2},
\end{equation}
where $\Delta \omega$ and $\Delta U$ are the change of atomic frequencies and gravitational potential between the measurements, respectively.
$R$ in (\ref{e:REEP}) is the relativistic factor which describes the deviation of the kinetic energy $E_K$ from the value given by the non-relativistic virial theorem (which states that $E_K=-E$, where $E$ is the total energy),
\begin{equation}\label{e:R}
R_{ab} = - \frac{E_{K,a} - E_{K,b}}{E_a-E_b}.
\end{equation}
The values of the factor $R$ are calculated in computer codes by varying the value of the kinetic energy operator in the Dirac equation
(see Ref.~\cite{EEP} for details). 

The results are very sensitive to the many-body effects which means that the effects should be treated very accurately or avoided. Otherwise the results are unstable. A good criterion for the reliability of the results is the achievement of the non-relativistic limit $R=1$.
This can be done by setting to zero the value of the fine structure constant $\alpha$ in the computer codes. It turns out that for complicated systems like those considered in present work the best results are obtained by simple estimations based on single-electron consideration. Namely, all clock transitions in Cu, Ag and Au can be considered as a $ns \rightarrow (n-1)d_{5/2}$ 
single-electron transitions ($n=4,5,6$ for Cu, Ag, Au respectively). Therefore, we just use single-electron energies of theses states in (\ref{e:R}).  We use the same approach for Hg$^+$ and Yb$^+$.

The results are presented in Table~\ref{t:q} together with results obtained earlier for other systems. Note, that the results for the transitions involving excitation from the $5d$ shell in Hg$^+$ and the $4f$ shell in Yb$^+$ are different from what was published before. 
The old calculations were based on a version of the CI method~\cite{CI1,CI2} which contained a fitting parameter responsible for correct energy interval between states of different configurations. It was assumed that this parameter does not change under variation of the kinetic energy operator. We believe that present results are more reliable because they are free from any assumptions and because they reproduce the non-relativistic limit $R=1$. Note that the values of $R$ for transitions in Yb and Yb$^+$, which do not involve excitation from the $4f$ shell are in good agreement with previous calculations ($R=1.12$ and $R=1.42$, see Table~\ref{t:q}). This means that present single-electron estimations work well and that accurate many-body calculations are possible for simple systems.

To study the LPI violation one needs to compare two clocks with different values of the relativistic factors $R$ (see formula (\ref{e:REEP})) over at least half of a year. Table~\ref{t:q} shows that there is a wide range of choices for such clock pairs. In particular, the Au clock is practically as good as the Hg$^+$ clock which was used before for this purpose in combination with the Al$^+$ clock~\cite{Hg+clock0,EEP}. The Cu and Ag clocks, which have $R$ values close to unity, can be used in combination with clocks  with large relativistic effects, e.g. Yb or Yb$^+$ clocks. 


\subsection{LLI violation}


The LLI violation term is a tensor operator
\begin{equation}\label{e:LLI}
\hat H_{\rm LLI} = -\frac{1}{6}C_0^{(2)} T_0^{(2)},
\end{equation}
where $C_0^{(2)}$ is unknown constant and the relativistic form of the $T_0^{(2)}$ operator is given by $T_0^{(2)} = c\gamma_0(\gamma^jp_j-3\gamma^3p_3)$.

To study the effect of the LLI violating term (\ref{e:LLI}) one needs long-lived atomic states with large value of the total electron angular momentum $J$, $J>1/2$. All clock states of Cu, Ag and Au satisfy this requirement. This term should cause a dependence of the atomic
frequencies on the apparatus orientation in space (e.g., due to Earth rotation). Interpretation of the measurements requires knowing the 
values of the reduced matrix elements of the operator $T_0^{(2)}$ for the clock states. We calculate these matrix elements using the CIPT
method to obtain wave functions and the RPA method to obtain effective operator for valence electrons. 

The results are presented in Table~\ref{t:q}. The results of earlier calculations for Yb~II~\cite{Yb+LLI} are also presented for comparison. In contrast to the search of the $\alpha$-variation and LPI violation, one clock state is sufficient for the search of the LLI violation. The comparison of frequencies is done for states with different projections of the total angular momentum $J$~\cite{Lia,Yb+LLI}. The large value of the matrix element is important but it is not the most important parameter, e.g. the lifetime of the metastable state is even more important (see, e.g.~\cite{Yb+LLI} for more discussion). The calculations for Cu and Au show that these systems are also suitable for the search of the LLI violation. Ag appears less well suited because its upper clock state has a significantly shorter lifetime.
 



\section{Conclusion}
We studied electric quadrupole transitions between ground and excited metastable states of Cu, Ag and Au and demonstrated that the transitions have all features of optical clock transitions.  Important systematic effects such as black-body radiation shift, Stark 
and Zeeman shifts, etc. are similar  to or smaller than in current top-performing optical clocks. On the other hand,  some of 
 the transitions are more sensitive to new physics beyond the standard model than the 
currently used  neutral-atom optical clocks,  or of complementary value. 
The studied effects included time variation of the fine structure constant, Local Lorentz Invariance violation and Local Position Invariance violation. 

\vskip 1in
\acknowledgements

This work was supported by the National Natural Science Foundation of China
(Grant No. 11874090) and the Australian Research Council.
V.A.D. would like to express special thanks to the Institute of Applied Physics and Computational
Mathematics in Beijing for its hospitality and support. 
This research includes computations using the computational cluster Katana supported by Research Technology Services at UNSW Sydney.



\appendix
\section{Stark and electric quadrupole shifts.}

Stark shift of the frequency of the transition between atomic states $a$ and $b$ due to interaction with residual static electric field $\varepsilon$ is
\begin{equation}\label{eq:St}
\delta \omega_{ab} = -\Delta \alpha_{ab}(0)\left(\frac{\varepsilon}{2}\right)^2,
\end{equation}
where $\Delta \alpha_{ab}(0)$ is the difference between static scalar polarizabilities of states $a$ and $b$. The shift is quadratic in electric field and usually small. It is further suppressed for considered clock transitions due to small difference in the polarizabilities (see Table~\ref{t:pol}).

The energy shift due to a gradient of a residual static electric field $\varepsilon$
is described by a corresponding term in the Hamiltonian
\begin{equation}\label{eq:HQ}
\hat H_Q = -\frac{1}{2}\hat Q\frac{\partial \varepsilon_z}{\partial z},
\end{equation}
where $\hat Q$ is the atomic quadrupole moment operator ($\hat Q = |e|r^2 Y_{2m}$, the same as for E2 transitions). 
The energy shift of a state with total angular momentum $J$ is proportional of the atomic quadrupole moment
of this state. It is defined as twice the expectation value of the $\hat Q$ operator in the stretched state
\begin{equation}\label{eq:Q}
Q_J = 2\langle J,J_z=J|\hat Q| J,J_z=J \rangle.
\end{equation}

Calculations using the CIPT method for wave functions and the RPA method for the operator 
give the values $Q_J=0.431$~a.u. for the $^2$D$_{5/2}$ clock state of Cu, $Q_J=0.296$~a.u. for the $^2$D$_{3/2}$ clock state of Cu, $Q_J=0.966$~a.u. for the clock state of Ag, and $Q_J=1.47$~a.u. for the clock state of Au. Quadrupole moments of the ground states of these atoms are zero due to small value of the total electron angular momentum ($J$=1/2).

Consider transitions between hyperfine structure (hfs) components of the ground and clock states with definite values of the total angular momentum $F$.
The quadrupole shift is proportional to $3F_z^2-F(F+1)$, where $F_z$ is the projection of $\mathbf{F}$.
For $F=3$ and $F_z=\pm 2$ this factor is zero and the quadrupole shift vanishes. Note that clock states with $F=3$ exist for all stable isotopes of all three considered atoms (see Table~\ref{t:isotopes}). {Using these states would lead to a linear Zeeman shift. It 
 cancels out by averaging over the transition frequencies to the states with $F_z=-2$ and $F_z=2$.}

\end{document}